# Pitch Perception of Complex Sounds: Nonlinearity Revisited [*]

D.L. Gonzalez, L. Morettini, F. Sportolari,
Sezione di Cinematografia Scientifica, Istituto Lamel, CNR
Bologna, Italy

O. Rosso,
Centro de Calculo, Universidad de Buenos Aires
Buenos Aires, Argentina

J.H.E. Cartwright and O. Piro
Departament de Física, Universitat de les Illes Balears
Palma de Mallorca, Spain

chao-dyn/9505001   5 May 95

## Abstract

The ability of the auditory system to perceive the fundamental frequency of a sound even when this frequency is removed from the stimulus is an interesting phenomenon related to the pitch of complex sounds. This capability is known as "residue" or "virtual pitch" perception and was first reported last century in the pioneering work of Seebeck. It is residue perception that allows one to listen to music with small transistor radios, which in general have a very poor and sometimes negligible response to low frequencies. The first attempt, due to Helmholtz, to explain the residue as a nonlinear effect in the ear considered it to originate from difference combination tones. However, later experiments have shown that the residue does not coincide with a difference combination tone. These results and the fact that dichotically presented signals also elicit residue perception have led to nonlinear theories being gradually abandoned in favour of central processor models. In this paper we use recent results from the theory of nonlinear dynamical systems to show that physical frequencies produced by generic nonlinear oscillators acted upon by two independent periodic excitations can reproduce with great precision most of the experimental data about the residue without resorting to any kind of central processing mechanism.

---

# 1 Introduction

From the beginning of acoustics research great efforts have been devoted to the elucidation of the mechanisms by means of which our auditive system can with astonishing performance analyse and discriminate between complex sounds. In particular, pitch perception has been a subject of great interest, most probably due to the key role played by pitch in music.

The first attempts to explain the pitch of complex sounds on a physical basis were made as early as the middle of the last century, just after Fourier methods were developed. The original approach, put forward by Ohm [1], considered pitch as a consequence of the ability of the auditory system to perform Fourier analysis on acoustical signals. In this view, a physical Fourier component of frequency $\omega_0$ is needed in the incoming stimulus in order to have a sensation of pitch matching that of a pure sinusoidal wave of the same frequency.

However, this approach quickly runs into contradictions. Almost contemporaneously with the work of Ohm, Seebeck [2] showed that if the fundamental frequency (or even the first few harmonics) is (are) removed from the spectrum of a periodic sound signal, the perceived pitch remains unchanged and matches the pitch of a sinusoidal sound with the frequency of the "missing fundamental". As some facts about the perception of the missing fundamental can be described quite naturally in terms of the stimulus periodicity, Seebeck proposed a "periodicity detection" theory for the pitch perception of complex sounds.

In this way was born an historical controversy between spectral and periodicity theories, which, in our opinion, is to date not completely resolved. The principal steps in this controversy are as follows:

— At the end of the last century, Helmholtz [3] reinforced Ohm's view, asserting that the ear acts as a rough Fourier analyser, and launched the hypothesis that this analysis is performed in the basilar membrane. Moreover, to explain the pitch of the missing fundamental, he proposed that a physical component at this frequency can be generated by the nonlinearities of the ear as a "difference combination tone".

— In the 1940's, with the aid of new electronic equipment, Schouten [4] realized a very well conceived experiment which demonstrated that the missing fundamental is not a difference combination tone. Schouten also elaborated a theory of pitch based on the periodicity properties of the non-resolved or "residue" components of the stimulus.

— In the 1960's von Bekesy [5] demonstrated experimentally that the hypothesis of Helmholtz is essentially correct, that is, the basilar membrane effectuates a rough Fourier analysis of the incoming stimulus.

— Later on, in the 1970's, central processor theories for pitch perception arose [6–8]. These were motivated by several failures of peripheral theories to explain psychophysical experiments and in the fact that dichotically presented stimuli also elicit residue perception.

The main drawbacks with peripheral theories are sensitivity to the phase relationship between partials in periodicity theories, and the impossibility of describing residue behaviour with difference combination tones in spectral theories.

In this paper, in the light of recent results in the theory of nonlinear dynamical systems, we revisit the possibility that ear nonlinearities produce physical components able to reproduce residue behaviour.

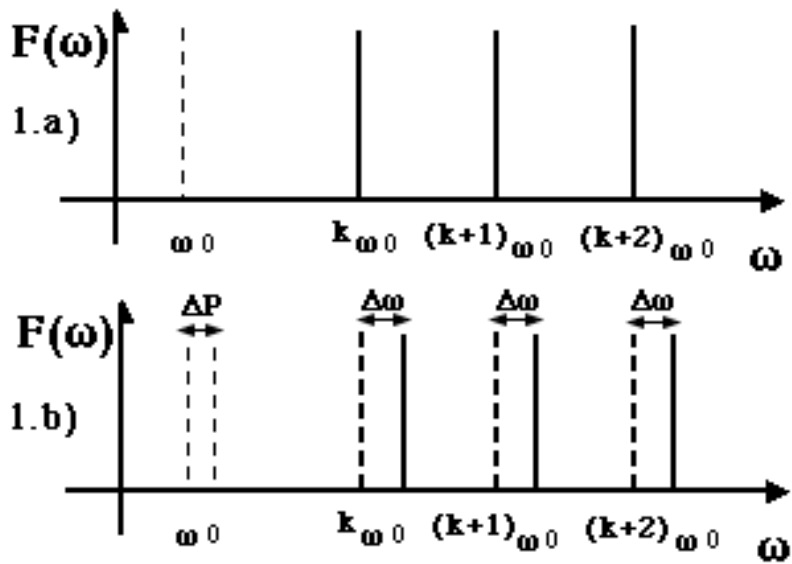

Figure 1: (a): Fourier spectrum of an harmonic complex sound. The partials are successive multiples $(k, k+1, \ldots)$ of some missing fundamental $\omega_0$ which determines the pitch of the complex. (b): Fourier spectrum of an anharmonic complex sound. The partials are obtained by a uniform shifting $\Delta\omega$ from the harmonic situation. Although the difference combination tones between successive partials remain unchanged and equal to the missing fundamental, the perceived pitch shifts by a quantity $\Delta P$ which depends linearly on $\Delta\omega$.

The paper is organized as follows: in the following section 2 we review the fundamental experimental facts about the residue. In section 3 we describe the behaviour of generic periodically forced nonlinear oscillators and we show how these results can be extended to the case of two independent periodic excitations. In section 4 we show that these results can be utilized to describe residue behaviour and finally in section 5 we discuss the implications of our work.

## 2 Residue Behaviour

Suppose that a periodic signal is presented to the ear. The pitch of the signal can be quantitatively well described by the frequency of the fundamental, say $\omega_0$. The number of harmonics and their relative amplitudes gives the timbral characteristics to the sound (the typical examples are musical sounds). Now suppose that the fundamental and perhaps some of the first few harmonics are removed (we call this series of partials, not necessarily multiples of a fundamental, a complex sound, or complex). Although the timbral sensation changes, the pitch of the complex remains unchanged and equal to the missing fundamental (see figure 1a). As we mentioned above, the first explanation of this phenomenon associated the residue with a difference combination tone. A difference combination tone arises from a nonlinear interaction of two pure tones and has a frequency given by the difference between the frequencies of the pure tones. For the case of an harmonic complex sound it is clear that the difference combination tone between two successive partials has the frequency of the missing fundamental i.e., $(n + 1)\omega_0 - n\omega_0 = \omega_0$. But if we now shift all the harmonics by the same amount $\Delta\omega$ (see figure 1b), the difference combination tone remains unchanged and the same should be true of the residue. This is basically the experiment realized by Schouten [4] with negative result: he found that the perceived pitch also shifts, showing a linear dependence on $\Delta\omega$. This phenomenon is known as first "pitch shift" effect and has been accurately measured in many independent experiments. In figure 2 we show a graph of the pitch shift for three different measurements. We can see that the slope of the lines decreases with increasing frequency of the central harmonic of the complex. A first attempt to model qualitatively the behaviour of the pitch shift shows that the slope depends roughly on the inverse of the harmonic number of the central partial of the complex, say $k$. However the change in slope is slightly but consistently larger than this (but smaller if we replace $k$ by $(k + 1)$). This behaviour is known as the second pitch shift effect. Finally, an enlargement of the spacing between partials while maintaining fixed the central frequency produces a decrease in the residue pitch. As this anomalous behaviour seems to be correlated with the second pitch shift effect it is usually included within it [4].

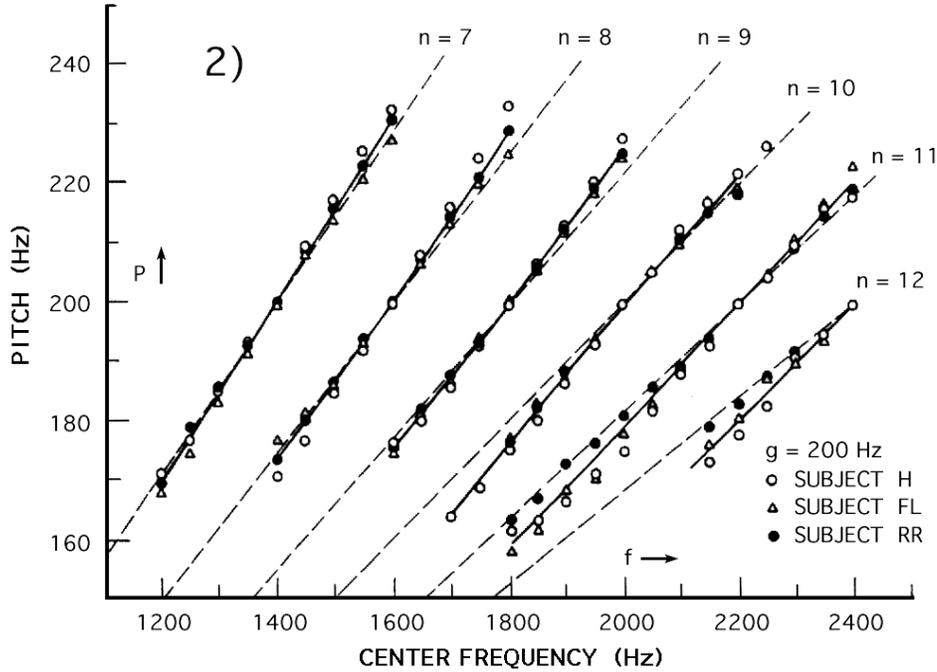

Figure 2: Pitch as a function of the central frequency $n = k + 1$ for a three-component complex tone $(k, k + 1, k + 2)$. Circles, triangles and dots represent data for three different subjects (from ref. [4]). The perceived pitch shifts linearly with the detuning. The dashed lines represents roughly the first pitch shift effect. Their slopes decrease as $1/n$ (see text).

## 3  Nonlinear Dynamics of Forced Oscillators

The basic idea in the following is to perform a qualitative modelling of the auditive system, identifying experimental data with structurally-stable behaviour of forced nonlinear oscillators. We can consider the ear as a nonlinear black box and the stimulus as a superposition of a variable number $n$ of purely sinusoidal waveforms.

With this aim we briefly review the fundamental dynamical features exhibited by a generic nonlinear oscillator in the case $n = 1$ and afterwards we show how these results can be extended to the case $n = 2$.

### 3.1.1  $n = 1$, synchronization

A periodically forced nonlinear oscillator can exhibit an extremely rich variety of responses. The most simple are periodic responses, known also as synchronized or phase-locked responses. In growing order of complexity we can encounter two-frequency quasiperiodic and chaotic responses. Due to space limitations we restrict our analysis to an heuristic presentation of some fundamental ideas about synchronization and quasiperiodicity (the interested reader can find additional details in ref. [9,10] and references included therein; a good review of the applications of chaos physics to acoustics is ref. [11]).

Usually, synchronization is detected as a rational ratio between the frequency of the external periodic force and the proper frequency of the oscillator (also between two proper frequencies in higher-dimensional autonomous systems). The first description of synchronization is due to Huygens [12], who observed that the pendulums of two clocks fixed on the same mounting after a time swung syncronously. In this case the two frequencies are equal and we say that we have a 1/1 synchronized response (where 1/1 stands for the frequency ratio). More complicated cases arise for an arbitrary rational frequency ratio. A beautiful example in nature is the 3/2 ratio between the orbital and rotation periods of the planet Mercury.

A typical forced oscillator, such as the forced van der Pol oscillator [13], shows an infinity of these phase-locked solutions depending on the values of the frequency and amplitude of the external force. For a constant value of the amplitude, the effective frequency ratio varies in a complicated manner with the external frequency, describing a kind of non-differentiable function, a fractal known as the devil's staircase (see figure 3). Every plateau in the graph corresponds to a particular phase-locked solution. The relative widths of the plateaux are locally organized in a hierarchical manner according to a number-theoretical property of the rationals which characterize them. We say that two rationals $p/q$ and $r/s$ are adjacents, if

$$|q.r - p.s| = 1. \tag{1}$$

Between adjacents we can define a Farey sum operation as follows

$$\frac{p}{q} \oplus \frac{r}{s} = \frac{p+r}{q+s}. \tag{2}$$

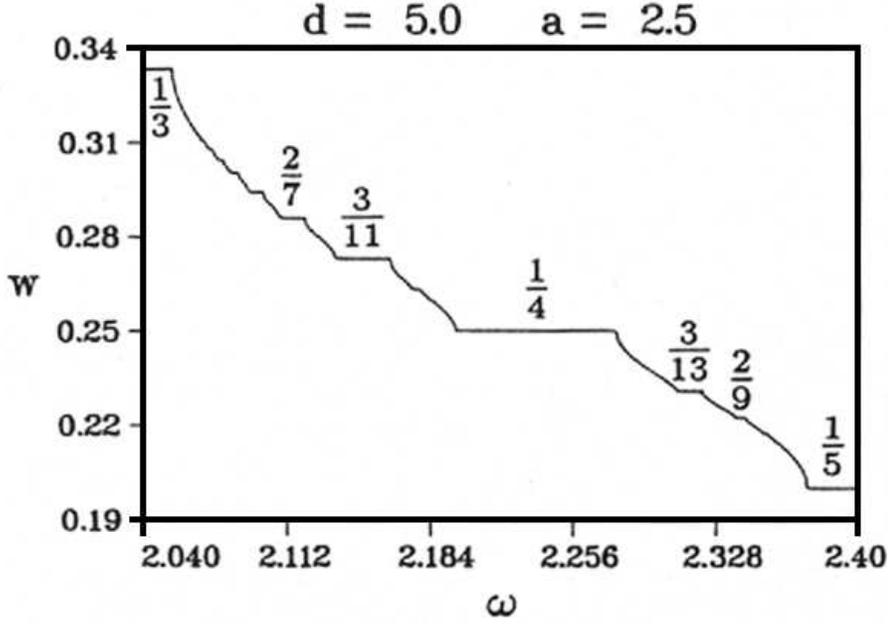

Figure 3: Devil's staircase for the forced van der Pol oscillator (from ref. [13]). Every rational plateau corresponds to a phase-locked solution, the denominator being the period of the response measured in periods of the external force.

The rational obtained is called the mediant of the two adjacents. The mediant gives the local hierarchy of the widths of the plateaux, determining the plateau with the greatest width between two plateaux characterized by adjacent rationals.

### 3.1.2  $n = 1$, quasiperiodicity

A quasiperiodic response can be expressed as a sum of periodic functions. The arguments are the linear combinations of a basic set of frequencies

$$\sum_{i=1}^{s} p_i - \omega_i, \qquad (3)$$

$s$ being the minimum integer for which the equation

$$\sum_{i=1}^{s} p_i \omega_i = 0 \qquad (4)$$

has no integer solutions others than the trivial one.

For $n = 1$ we can have quasiperiodic responses of order $s = 2$. Observe that difference combination tones, such as $(\omega_2 - \omega_1)$ for example, are included in this class. As we mentioned above, difference combination tones are not adequate for describing the residue. Moreover, two-frequency quasiperiodic responses are structurally unstable in the sense that small perturbations of the system destroy them. Consequently, having in mind the search for structurally stable responses able to reproduce residue behaviour, we increase the dimensionality of the system to $n = 2$, that is, a nonlinear oscillator forced with two independent external periodic forces.

### 3.2.1 $n = 2$, synchronization

When the external frequency ratio is irrational, periodic solutions cannot exist. When this ratio is rational they are destroyed by small perturbations of the external forces and consequently are not useful for a description of robust behaviour of the auditive system.

### 3.2.2 $n = 2$, three-frequency resonances

In the case $n = 2$ we can have three-frequency quasiperiodic responses. However, an important result in the theory of dynamical systems, the theorem of Ruelle–Takens–Newhouse [14], asserts that three-frequency quasiperiodic responses are structurally unstable, and thus of no interest for our purposes.

Another possibility (the last if we exclude chaotic solutions) is three-frequency resonant responses. Three-frequency resonances cannot be expressed as a linear superposition for $s = 2$, but are not truly three-frequency quasiperiodic in the sense that eq.4 has nontrivial integer solutions for $s = 3$.

In a previous work we showed that these responses are structurally stable for two particular systems (an electronic oscillator and an exactly solvable system of nonlinear differential equations). Moreover, we found that they are hierarchically organized in a similar way to the case of phase-locked responses for $n = 1$. In the same work, we extended the number-theoretical approach briefly described in section 3.1.1 to the case of three-frequency resonances. In the following we review succinctly the main results, the details being beyond the scope of this article (see ref. [15]). If $\omega_1$ and $\omega_2$ are the two external frequencies and their frequency ratio can be approximated by a rational $p/q$, we can define a generalized Farey sum operation between two fractions of real numbers, say $\omega_i/r_i$ and $\omega_j/r_j$, modifying the adjacency condition of eq.1: we say that the two fractions are adjacents if they satisfy

$$|\omega_i.r_j - \omega_j.r_i| = |\omega_1.q - \omega_2.p|. \qquad (5)$$

Now we can define the generalized mediant between these adjacents as

$$\frac{\omega_i}{r_i} \oplus \frac{\omega_j}{r_j} = \frac{\omega_i + \omega_j}{r_i + r_j}. \qquad (6)$$

Starting with $\omega_2/q$ and $\omega_1/p$ and recursively applying eq.6 we can construct a hierarchical structure of three-frequency resonances. The first steps of this structure together with experimental results obtained with an electronic oscillator are shown in figure 4.

## 4 A Nonlinear Theory for the Residue

We consider now the problem of the residue as a multiperiodic forcing of a generic nonlinear system which roughly represents the auditive periphery. The more simple case for the stimulus is a complex sound consisting of only two partials, say $k$ and $k + 1$, lying in the vicinity of successive multiples of some missing fundamental $\omega_0$.

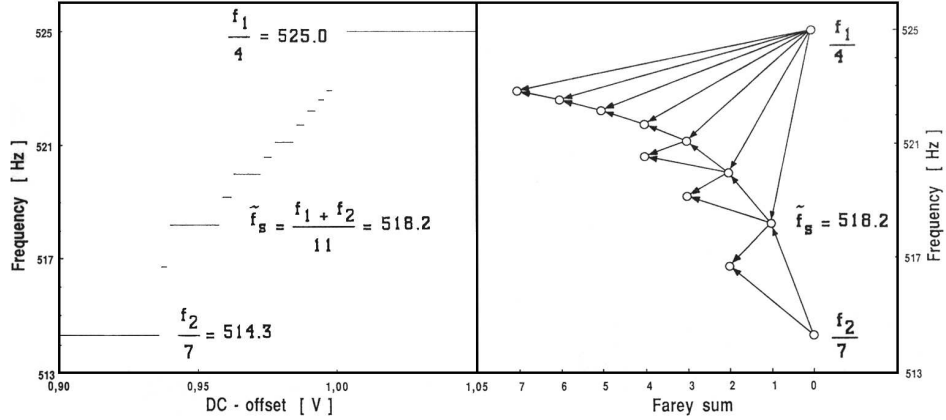

Figure 4: On the left side we have plotted the third frequency ($f_i = 2\pi\omega_i$) of a three-frequency resonance vs. a control parameter (the DC offset of one of the external forces). The three-frequency resonances are obtained as structurally-stable responses of a nonlinear electronic oscillator driven by two independent external periodic forces. On the right side we can see the frequency values predicted by means of the generalized Farey sum operation. The external frequencies are fixed at 2100 Hz and 3600 Hz.

Now we search for structurally stable solutions which could be associated with the residue. As we have seen, periodic solutions are structurally unstable to perturbations of the external stimulus. Two-frequency quasiperiodic solutions (difference combination tones) are unable to reproduce residue behaviour. Three-frequency quasiperiodic solutions are structurally unstable to perturbation of the system's parameters (Ruelle–Takens–Newhouse theorem). There remain only two possibilities, three-frequency resonant solutions and chaotic ones. Bearing in mind the results of section 3.2.2, we propose that the residue is associated with the third frequency in a three-frequency resonance formed by a frequency generated in the auditive system itself (in the vicinity of the missing fundamental $\omega_0$) and two external frequencies (in the vicinity of $k\omega_0$ and $(k + 1)\omega_0$, respectively).

The vicinity of the external frequencies to successive multiples of some missing fundamental ensures that $k/(k + 1)$ is a good rational approximation to their frequency ratio. Consequently, from the results of section 3.2.2, $\omega_2/(k + 1)$ and $\omega_1/k$ are adjacents. With the aid of eq.6 we obtain the value of the third frequency in the three-frequency resonance of greatest width between them

$$\frac{\omega_1}{k} \oplus \frac{\omega_2}{k+1} = \frac{\omega_1 + \omega_2}{2k+1}. \tag{7}$$

Since the external frequencies can be written as (equal detuning):

$$\omega_1 = k\omega_0 + \Delta\omega, \qquad \omega_2 = (k+1)\omega_0 + \Delta\omega, \tag{8}$$

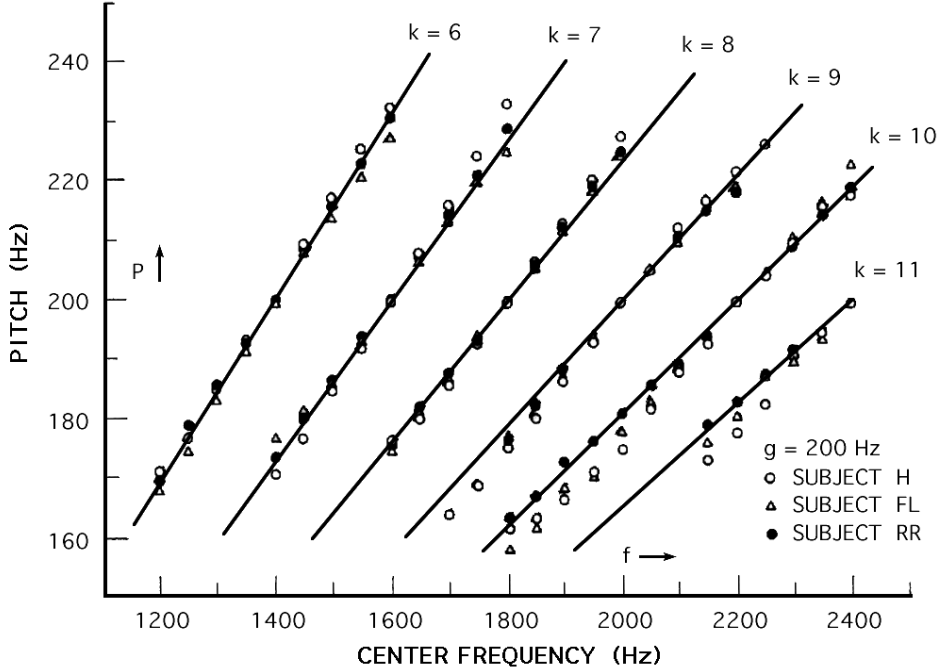

Figure 5: Plot of the predicted pitch shift effect (eq.9) on the data of figure 2.

the shift of the third frequency with respect to the missing fundamental is:

$$\Delta P = \frac{2\Delta\omega}{2k+1}. \qquad (9)$$

This equation gives a linear dependence of the shift on the detuning $\Delta\omega$, in accordance with the first pitch shift effect (see figure 2). The predicted slope is $1/(k+1/2)$, just in the middle between $1/k$ and $1/(k+1)$. In figure 5 we have superimposed the behaviour of the corresponding three-frequency resonances on the data of figure 2. The fit is very good, explaining the first aspect of the second pitch shift effect (section 2). The second aspect can be interpreted as follows: the term $2\Delta\omega$ in eq.9 arises from two equal contributions $\Delta\omega$ obtained by means of a uniform shift in the two forcing frequencies. If now, maintaining $\omega_2$ fixed we increase the distance to $\omega_1$ (we enlarge the spacing between successive partials) the first contribution remains constant and equal to $\Delta\omega$ while the second diminishes, determining a decrease in the third frequency of the resonance and thus in the residue (see eq.9).

## 5 Discussion

We have shown that physical Fourier components, which are generated as structurally-stable responses of forced nonlinear oscillators, can adequately describe residue behaviour, that is, the pitch of complex sounds. Some questions arise quite naturally if our hypothesis is to be confirmed. The first in importance is: where are three-frequency resonances generated?

Two distinct experiments show that there do not exist physical components in the region of the basilar membrane giving a maximal response at the frequency of the residue. In fact, a test sinusoidal tone with frequency in the vicinity of the residue does not produce beats [4]. Also a masking signal with a spectrum centred on the residue frequency is unable to mask it [16]. Masking is possible only when the masking signals are centred on the primary tones. These results strongly suggest that if there exists a physical cochlear response, it must be located in some region excited by the primary tones. In our view, this place should be nearer the stapes than the region of optimal response to the lowest frequency in the spectrum; at some intermediate point where the effective forcings due to the lowest and near-lowest components in the stimulus are comparable. Moreover, nonlinear oscillations need some kind of feedback. Outer hair cells are good candidates for this. Outer hair cells could be the elements where three-frequency resonances are produced. This could also explain the mechanism by which outer hair cells, tailored for very low frequency responses, could interact dynamically with the higher-frequency components of the stimulus (the third frequency is always lower than the frequencies of the forcing terms). Outer hair cells could furnish a set of "pitch candidates" for some more central regions of the auditory system. In turn, these regions could then choose a particular candidate on the basis of frequency and amplitude rules and feed this temporal information back to the peripheral system in order to perform more complex processing of the acoustic signal. Some evidence of physical responses of the cochlea at subharmonic frequencies has been given by laser-velocimetry experiments [17]. Evidence of some kind of complex subharmonic processing in the cochlea can be extracted from experiments on interaction of spontaneous oto-acoustic emissions with two-tone external sounds [18]. Moreover, auditory-evoked potentials recorded from the scalp have shown the existence of physical components at the residue frequency [19].

The residue is not merely an acoustical curiosity. Its importance is shown in our ability to listen to music in a small transistor radio with negligible response to low frequencies. In fact, residue and musical perception seems to be profoundly correlated. Residue perception is probably at the base of musical harmony, and a parallel with the concept of fundamental bass, first proposed by Rameau [20], can be hypothesized (we should remark that we refer to "musical consonance", not "psychoacoustic consonance" which is related to the concept of critical band [8]). Moreover, the residue seems to play an important role in speech intelligibility. Hearing aids which furnish fundamental frequency information produce better scores in profoundly hearing impaired subjects than amplification [21]. It is clear that an improvement in the knowledge about the basic mechanisms involved in pitch perception may allow a similar improvement in hearing aids through the implementation of analog compensation processing (some kind of intelligent amplification) of the acoustical signals.